\documentclass[]{interact}

\usepackage{epstopdf}% To incorporate .eps illustrations using PDFLaTeX, etc.
\usepackage[caption=false]{subfig}% Support for small, `sub' figures and tables

\usepackage[numbers,sort&compress]{natbib}% Citation support using natbib.sty
\bibpunct[, ]{[}{]}{,}{n}{,}{,}% Citation support using natbib.sty
\renewcommand\bibfont{\fontsize{10}{12}\selectfont}% Bibliography support using natbib.sty
\makeatletter% @ becomes a letter
\def\NAT@def@citea{\def\@citea{\NAT@separator}}% Suppress spaces between citations using natbib.sty
\makeatother% @ becomes a symbol again

\theoremstyle{plain}% Theorem-like structures provided by amsthm.sty

\theoremstyle{definition}

\theoremstyle{remark}

\def\ltsima{$\; \buildrel < \over \sim \;$}
\def\simlt{\lower.5ex\hbox{\ltsima}} \def\gtsima{$\; \buildrel > \over
\sim \;$} \def\simgt{\lower.5ex\hbox{\gtsima}}

\begin{document}

%\articletype{ARTICLE TEMPLATE}% Specify the article type or omit as appropriate

\title{Dark Energy: is it `just' Einstein's  Cosmological Constant $\Lambda$?}

\author{
\name{Ofer Lahav\textsuperscript{a}\thanks{CONTACT Ofer Lahav. Email:  o.lahav@ucl.ac.uk}}
\affil{\textsuperscript{a} Department of Physics \& Astronomy, University College London, Gower Street, London WC1E 6BT, UK }}

\maketitle

\begin{abstract}
The Cosmological Constant $\Lambda$, a concept introduced by Einstein in 1917, has been with us ever  since in different variants and incarnations, including the broader concept of 
`Dark Energy'.
Current observations are consistent  with a value of  $\Lambda$ corresponding to about present-epoch 70\% of the critical density of the Universe.
This is  causing the speeding up  (acceleration) of the expansion of the Universe over the past 6 billion years, a discovery recognised by the 2011 Nobel Prize  in Physics. Coupled with the flatness of the Universe and the amount of 30\% matter (5\% baryonic and 25\%  `Cold Dark Matter'),
 this forms the so-called `$\Lambda$CDM standard model', which has survived many observational tests  over about 30 years.
However,  there are currently indications of  inconsistencies (`tensions' ) within $\Lambda$CDM on different values of the Hubble Constant and the  clumpiness factor. Also, time variation of Dark Energy and slight deviations from General Relativity are not ruled out yet. Several grand projects are underway to test $\Lambda$CDM further and to estimate the cosmological parameters to sub-percent level.
If $\Lambda$CDM will  remain  the standard model, then the ball is back in the theoreticians' court, to explain the physical meaning of $\Lambda$.  Is $\Lambda$  an alteration to the geometry of the Universe, or the energy of the vacuum?  Or maybe it is something different, that manifests a  yet unknown higher-level theory? 
\end{abstract}


%---------------------------------------------------------
\section{Introduction}
\begin{figure}

\includegraphics[width=0.9\textwidth]{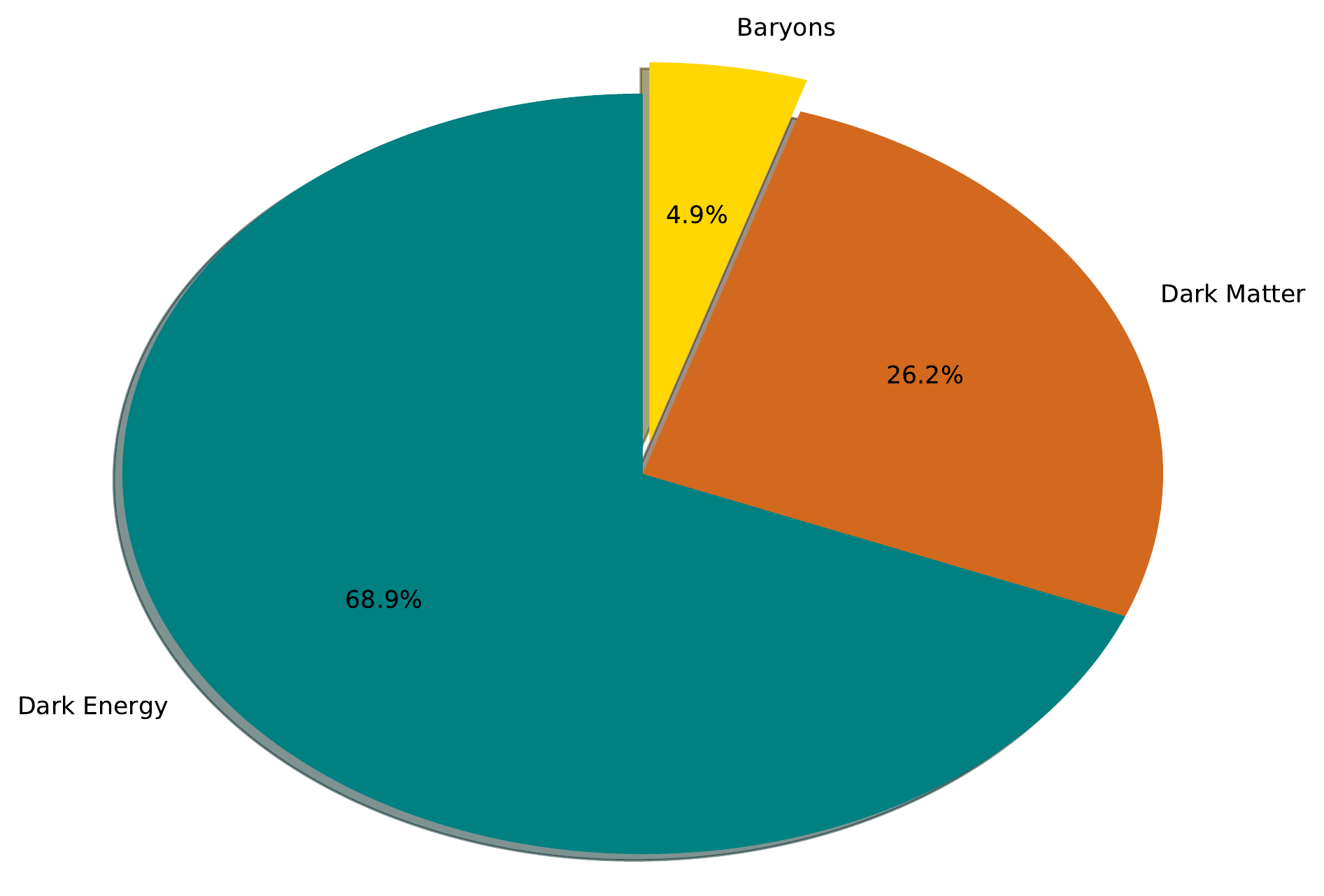}

\caption{ The relative amounts of the different constituents of the Universe: Baryonic  (ordinary) matter, Dark Matter and Dark Energy.   The plot is based on the results from the Planck collaboration \cite{Planck2018I}.
}
\label{Planck}
\end{figure}

It is well established that the Universe started in a Big Bang about 13.8 billion years ago. The expansion of the Universe is a natural  outcome of Einstein's theory General Relativity,
for particular the mass/energy contents of the Universe, but it is still a mystery what the Universe is  actually made of. 
The expansion can be described by a single function of time, the cosmic scale factor, $a(t)$, which determines how the distances between galaxies change in time. It can also be expressed as $a = 1/(1+z)$,  where $z$ is the cosmological redshift.   
The rate of expansion is defined as $H(t) = {\dot a}/a$ (dot indicates derivative with respect to time),  where the present-epoch value is the Hubble Constant $H_0$. Recent observations indicate that the universe is not only expanding - it is also accelerating at the present epoch, ${\ddot a} > 0$. But what is accelerating the Universe?
The composition of the Universe, agreed by most astronomers, is a rather bizarre mixture: baryonic (ordinary) matter, Cold Dark Matter, massive neutrinos and Dark Energy. 
Dark Energy is a broad term, which includes  the Cosmological Constant  $\Lambda$, a concept  born in 1917, in Einstein's famous article \cite{Einstein1917}.  It is accepted that about 70\% of the present-epoch Universe is made of `Dark Energy', a mysterious substance  which causes the cosmic acceleration.
 A further 25\% of the Universe is made from invisible `Cold Dark Matter' that can only be detected through its gravitational effects, with the ordinary atomic matter making up the remaining 5\%, and an additional tiny contribution from massive neutrinos, with its exact value still unknown 
This composition is illustrated in Figure \ref{Planck}, based on  the Planck Collaboration results \cite{Planck2018I}.
This ``$\Lambda$ + Cold Dark Matter"($\Lambda$CDM) paradigm and its extensions pose fundamental questions about the origins of the Universe. If Dark Matter and Dark Energy truly exist, we must understand their nature. Alternatively, General Relativity and related assumptions may need radical modifications. These topics have been flagged as key problems by researchers and by advisory panels around the world,  and significant funding has been allocated towards large surveys of Dark Energy. Commonly, Dark Energy is quantified by an equation of state parameter  $w$ (the ratio of pressure to density, see below). The case where $w = - 1$ corresponds to Einstein's  $\Lambda$ in General Relativity, but in principle $w$ may vary with cosmic time. Essentially, $w$ affects both the geometry of the Universe and the growth rate of structures. These effects can be observed via a range of cosmological probes, including the CMB, Supernovae Type Ia,  galaxy clustering, clusters of galaxies, and weak gravitational lensing. The 
 Type Ia Supernova surveys \cite{Riess1998, Perl1999} revealed that our Universe is not only expanding but is also accelerating in its expansion. 
The 2011 Nobel Prize in Physics was awarded for this remarkable discovery. Evidence for cosmic acceleration was actually noted even earlier in the 1990s, when galaxy clustering measurements \cite{Efstathiou1990} indicated a low matter density parameter. This suggested the possibility of a Cosmological Constant  
when combined with the assumption that space is `flat' (e.g. two light beams would travel in parallel lines).
Flatness (zero curvature) is predicted by Inflation, a theory of exponential expansion in the early Universe, in a tiny fraction of a second just after the Big Bang.
The flatness of the universe was later confirmed by CMB  anisotropy measurements. 
%-------------------------------------------
%---------------------------------------------

We start this review by providing some background and history on  $\Lambda$ in General Relativity and on the concept of Dark Energy and  a Newtonian version for the $\Lambda$ term (Section 2).
Then we summarise  the modern probes of Dark Energy (Section 3) and the landscape of galaxy surveys (Section 4), before presenting results from the Dark Energy Survey (Section 5) and discussing tension in the Hubble Constant (Section 6). We close by  considering the outlook for Dark Energy research (Section 7). 

\section{Background: a brief history of Dark Energy}
\label{sec:2}
%--------------------
\begin{figure}
\includegraphics[width=0.9\textwidth]{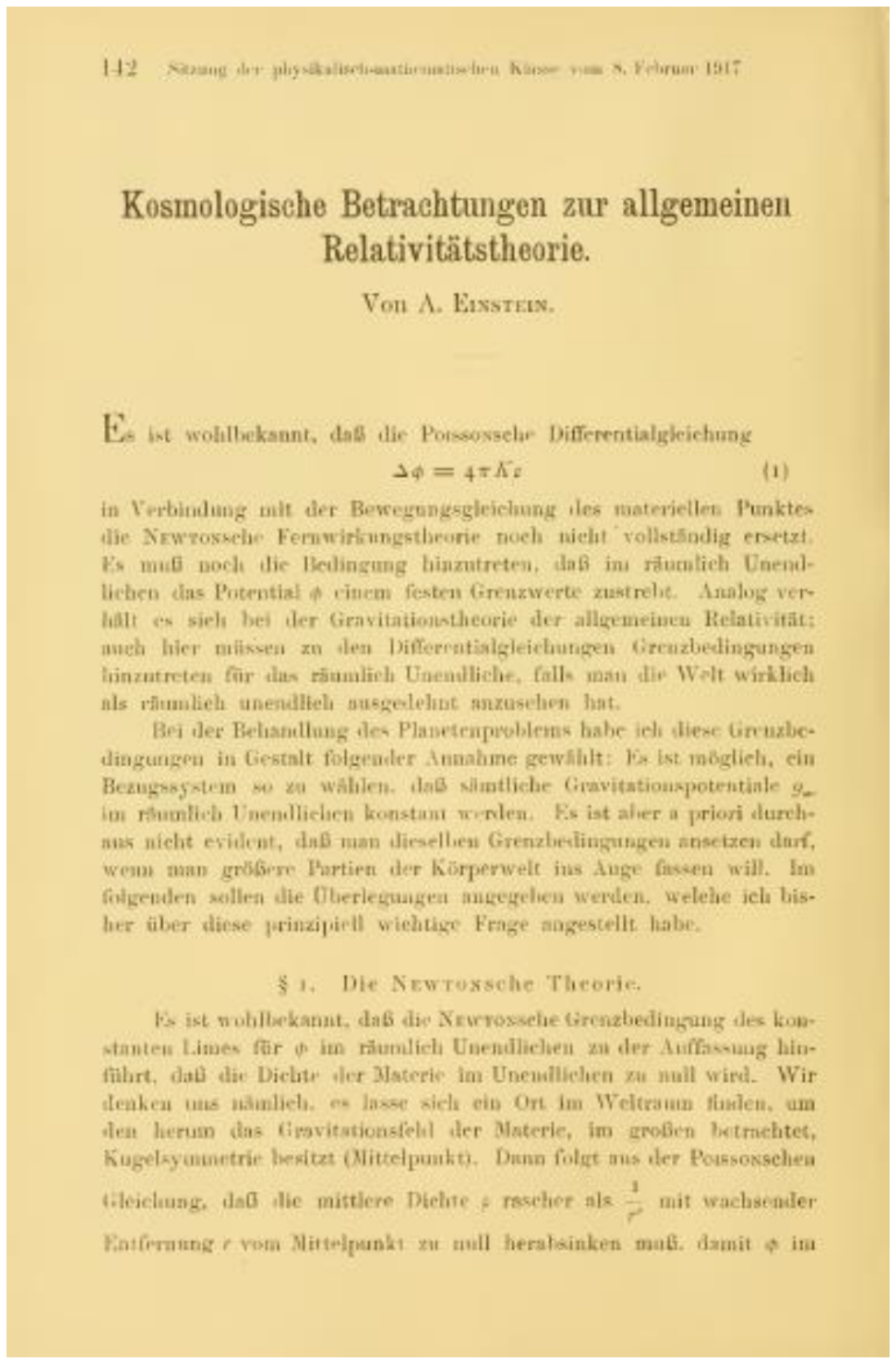}
\caption{The first page of Einstein's 1917 paper \cite{Einstein1917} on ``Cosmological considerations in the General Theory of Relativity" (in German).
}
\label{Einstein1917}
\end{figure}
%-------------------------------------------

%--------------------
\begin{figure}
%from the Guardian May 2019
\includegraphics[width=0.9\textwidth]{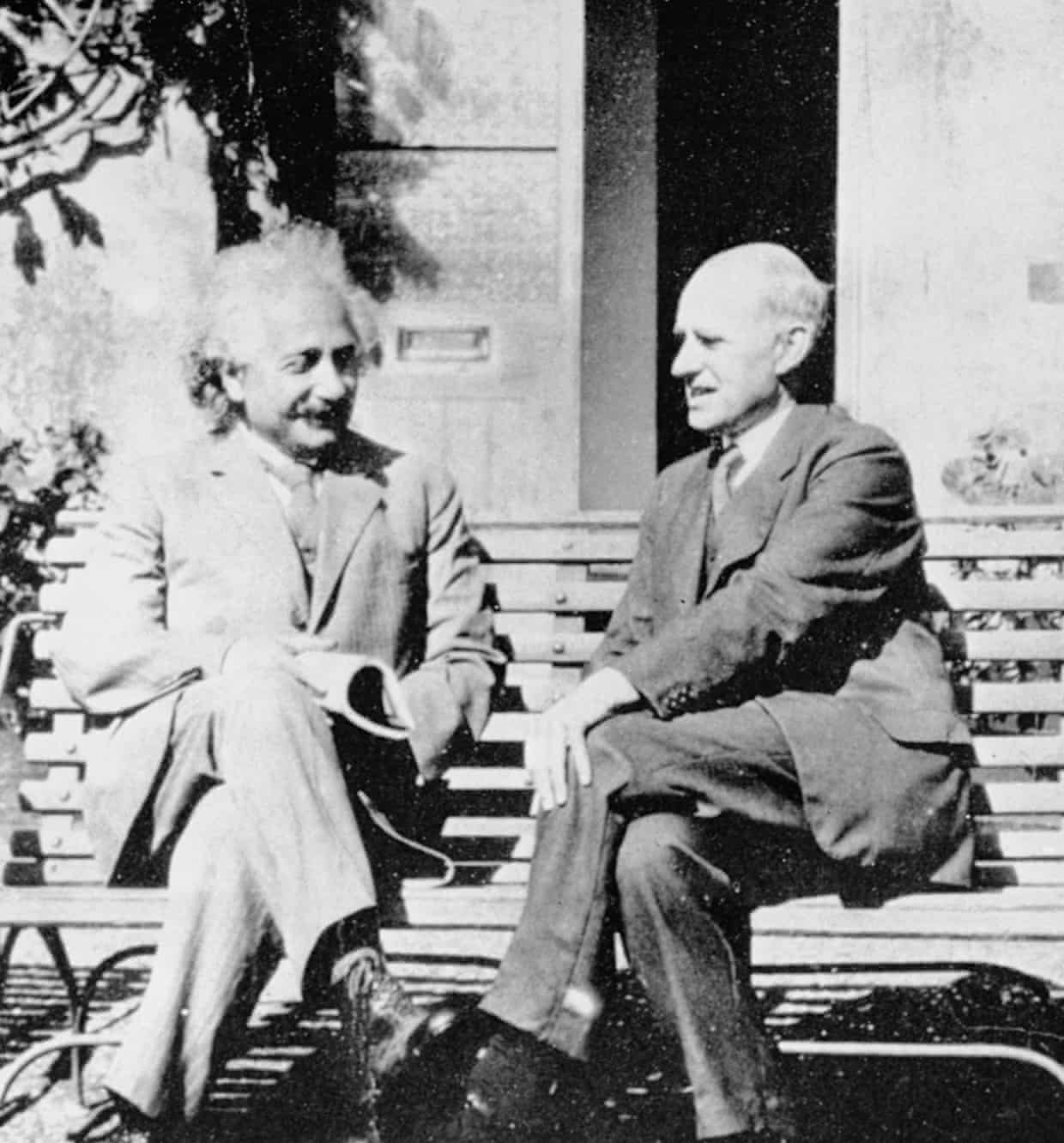}

\caption{Arthur Eddington (right) and Albert Einstein at the Cambridge Observatory (now part of  the Institute of Astronomy), in 1930.   Einstein introduced the Cosmological Constant $\Lambda$ in 1917, but  then deserted it, while Eddington favoured keeping $\Lambda$  as part of General Relativity. Credit: Royal Astronomical Society/Science Photo Library } 
\label{Einstein_Eddington}
\end{figure}
%-------------------------------------------

Over a century ago
Einstein added  the  Cosmological Constant $\Lambda$ to his equations \cite{Einstein1917}  (see Figure \ref{Einstein1917}).
It is a complicated equation, but we can think of it in a symbolic way as:
 \begin{equation}
 G +  \Lambda = T,
\end{equation}
%His full equation is then: 
%\begin{equation}
%R_{\mu \nu} - {1\over 2}  R g_{\mu \nu} + \Lambda g_{\mu \nu}  = {{8 \pi G} \over {c^4}} T_{\mu \nu} \;.
%\end{equation}
%Here, $R_{\mu \nu} $ and  $R$ are the Ricci tensor and scalar curvature of spacetime, respectively, 
%$g_{\mu \nu}$ is the metric tensor, $T_{\mu \nu}$ is the energy-momentum tensor of all matter and radiation, $G$ is Newton's gravitational constant, and $c$ is the speed of light. 
where $G$ represents the gravity in space-time, and $T$ (called the `energy-momentum tensor') describes  the contents of the Universe.
One way to interpret this equation is in the words of John Archibald Wheeler: `Matter tells spacetime how to curve; curved spacetime tells matter how to move.' 
%We note that Einstein put  $\Lambda$  on the left hand side, as a 'modification' to the the curvature term.
Einstein added $\Lambda$ primarily as a way of generating a static universe, which neither expands nor contracts.
With Edwin Hubble's observations of the expansion of the universe a few years later it seemed to Einstein that the $\Lambda$ term was not needed anymore,  
and he blamed himself that inventing $\Lambda$ was the `blunder of his life'\footnote{For a historical review of Einstein's 1917 paper see  \cite{OR2017}.} .  
One may argue that Einstein  `missed opportunities' in a number of ways: to predict the expansion/contraction of the Universe without $\Lambda$,
to notice that his static solution is unstable, so  a Universe with $ \Lambda$  would also expand/contract, 
and to view  $\Lambda$  as a free parameter, which could also lead for example to an accelerating Universe.
But this  1917 study was probably the first paper on relativistic cosmology,
and the  first  time the cosmological constant $\Lambda$ was proposed. Dark Energy, an extension of the $\Lambda$, is  now the focus of projects worth billions of dollars, with hundreds of scientists spending significant parts of their careers on this problem. 

While Einstein abandoned $\Lambda$,  his  friend Arthur Stanley Eddington (Plumian Professor at Cambridge; see Figure \ref{Einstein_Eddington} ) was actually very keen on the Cosmological Constant,  considering it as the possible source of expansion. In his popular book `The Expanding Universe' \cite{Eddington1920} he stated: ``I am a detective in search of a criminal - the Cosmological Constant. 
 I know he exists, but I do not know his appearance, for instance I do not know if he is a little man or a tall man....".
 
%The landscape of universes with different values of $\Lambda$ and matter density is illustrated in Figure 3.
The big conceptual question is if $\Lambda$ should be on the left hand side of equation (1), as part of the curvature (as originally introduced by Einstein), or on the right hand side, as part of the energy-momentum tensor $T$,
%$T_{\mu \nu}$, 
for example associated with the vacuum energy\footnote{It seems non-intuitive that the vacuum contains energy. But recall in Quantum Mechanics even the ground state of a simple harmonic oscillator has energy,  $E=\frac{1}{2}h \nu$. Summing up the contributions from many harmonics oscillators (up to a certain energy cutoff) gives rise to vacuum energy.
Vacuum energy can also be viewed in terms of `virtual particles' (also known as `vacuum fluctuations'), which are created and destroyed out of the vacuum. Vacuum energy has been verified experimentally in the lab by the Casimir effect.} $\Lambda = 8 \pi  G \rho_{vac}/c^2$. 
In fact, oddly,   Quantum field theory predicts for the amount of vacuum energy  $10^{120}$ times the observed value; that is a challenging `Cosmological Constant problem' by itself (see \cite{Weinberg89}).
If Dark Energy is the energy of the vacuum, then (unlike matter and radiation) its energy density doesn't dilute as the Universe expands. 
%On the other hand, Dark Energy could be associated with, other entities (e.g. a time-varying scalar field), in which case its energy density would evolve in time. 
More generally this time evolution is determined by the equation of state parameter $w$ of Dark Energy, defined as the ratio of its pressure $p$ to its energy density $\rho$, 
\begin{equation}
w = p/\rho c^2~.
\end{equation}
The Dark Energy density varies with the scale factor $a(t)$   as 
\begin{equation}
\rho \propto a^{-3(1+w)}~.
\end{equation}
For vacuum energy, the equation of state parameter is $w=-1$  and the density does not change with time, essentially the $\Lambda$ case  (see Figure 4). The density  of matter 
(the sum of ordinary matter  and Cold Dark Matter) varies like $\rho_{\rm m} \propto a^{-3}$, so 
the matter and $\Lambda$ densities were equal at  
\begin{equation} 
z_{\rm eq} = {(\Omega_{\Lambda}/\Omega_{\rm m})}^{1/3} - 1   \approx  0.30\;, 
\end{equation}
where the density parameter for $\Lambda$ is normalised by the Hubble Constant $H_0$, 
$\Omega_{\Lambda} \equiv \Lambda/(3 H_{0}^{2}) =0.7$, 
and the matter density $\rho_{\rm m}$ is scaled by the critical density\footnote{The critical density is defined as $\rho_{\rm c} \equiv 3H_0^2/8 \pi G$.  For Hubble constant 
$H_0 = 70 \;{\rm km} \, {\rm s}^{-1} \, {\rm Mpc}^{-1}$ 
it is 
$\rho_{\rm c} \approx 9.2 \times 10^{-27}$  kg m$^{-3}$.  This is a very low density, just over five protons per cubic metre of space.} 
$\Omega_{\rm m} \equiv \rho_{\rm m}/\rho_{\rm c}= 0.3$.
The  acceleration actually started earlier, at 
\begin{equation}
z_{\rm acc} = {(2 \Omega_{\Lambda}/\Omega_{\rm m})}^{1/3} - 1  \approx  0.64 \;.
\end{equation}
%(about 6.0 billion years ago). 
We note that 
$(1+z_{\rm acc})/(1+z_{\rm eq}) = 2^{1/3} \approx 1.26$, 
regardless of the values  of   $\Omega_{\Lambda} $ and $\Omega_{\rm m} $.
%In other models for Dark Energy, $w$ may differ from $-1$ and itself may evolve in time, leading to a time-dependent Dark Energy density.

In the Newtonian limit of General Relativity the equation of motion looks familiar:
\begin{equation}
{ d^2 r \over dt^2 } = - {{GM} \over {r^2} } + {c^2 \over 3} \Lambda r \;.
\end{equation}
In addition to the  famous inverse square law,
the second term is a linear force, that surprisingly was already discussed by Newton in Principia  (see  e.g. \cite{CL2008} for review).
An intuitive Newtonian way to think about the $\Lambda$ term is as a repulsive linear force,  opposing the inverse squared gravitational force.
It is interesting that such a force can be noticeable on the Mpc scale\footnote{Mpc stands for Megaparsec, which is a distance of million parsecs, equivalent to about 3.26 million light years,
or $3.09 \times 10^{19}$ kilometers.}. For example,
the $\Lambda$-force affects the acceleration of the Milky Way and Andromeda towards each other
  (see e.g. \cite{BT2008, PLH2013, McLeod2016}). 
For further discussion on Dark Energy and Cosmological Parameters see reviews in \cite{Weinberg2019, Lahav_Liddle2019, Lahav1991, CL2010}.

%--------------------
\begin{figure}
\includegraphics[width=0.7\textwidth]{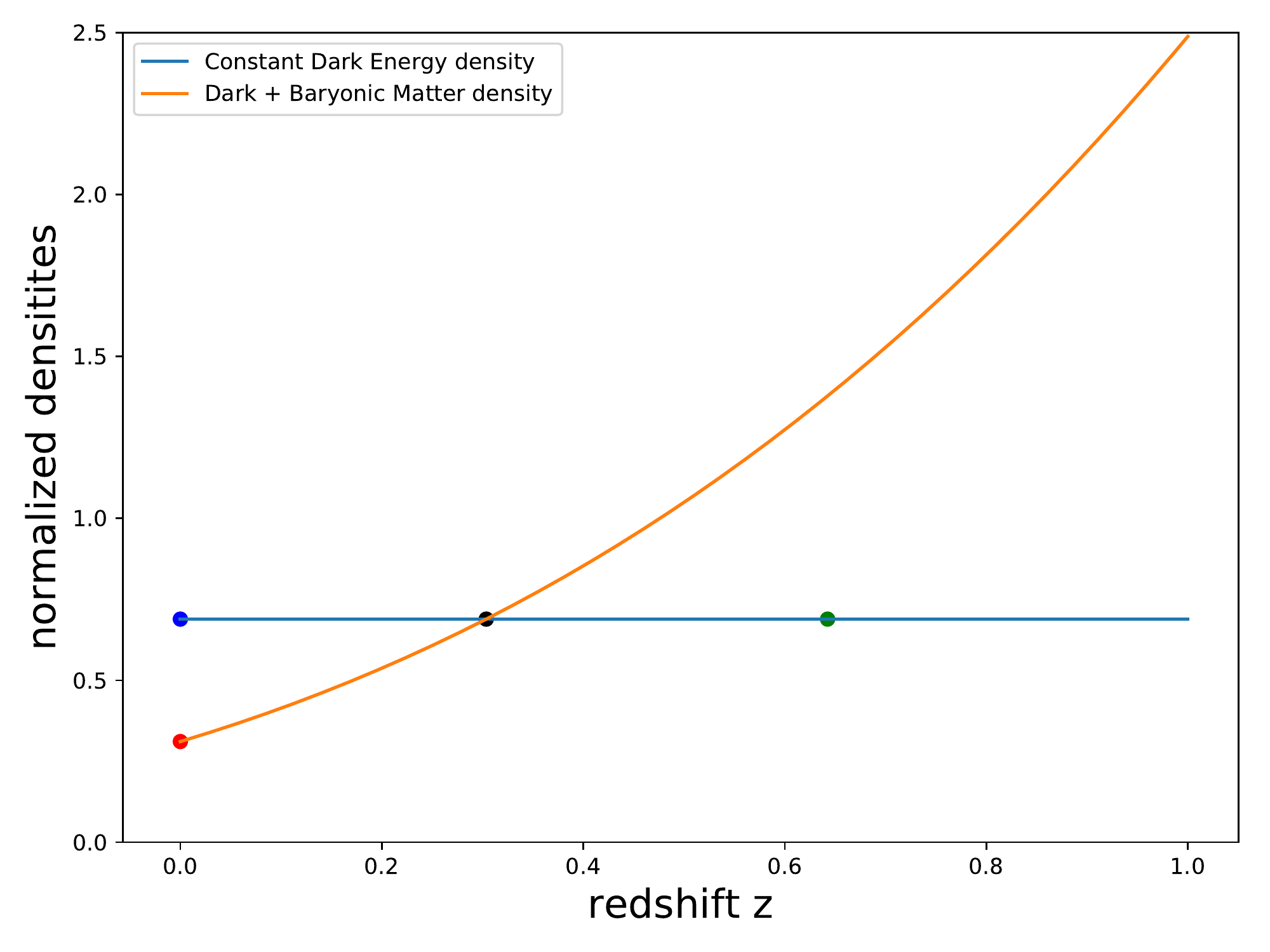}
\caption{The variation of densities with cosmic time:   matter $\rho_{\rm m}  \propto a^{-3}$  and a constant Dark Energy 
$ \rho_{\Lambda}$ corresponding to $w=-1$ . 
The blue and orange dots at $z=0$ indicate the present approximately observed density parameters, $\Omega_{\Lambda} =0.7$ and $\Omega_{\rm m} = 0.3$.
As discussed in the text, the crossing of the two curves, when the two densities were equal, happened at 
$z_{\rm eq} 
 \approx  0.30 $ (about 3.5 billion years ago, for Hubble Constant $H_0 = 70$ km s$^{-1}$Mpc), 
while acceleration started earlier (marked by the rightmost dot), at 
$z_{\rm acc} 
 \approx 0.64 $
(about 6 billion years ago). For reference, our Solar System formed about 4.5 billion years ago.}
 \label{DE_densities}
\end{figure}
%-------------------------------------------

\section{Primary probes of Dark Energy}

%-------------------
\begin{figure}
\includegraphics[width=0.8\textwidth]{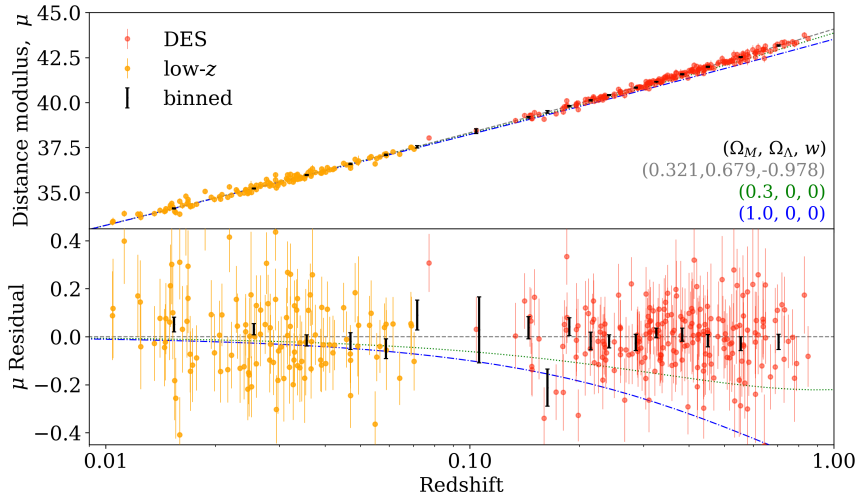}
\caption{Supernovae Type Ia as standard candles:
Hubble diagram for the Dark Energy Survey (DES) supernova (SN) 3YR sample (red) and other low-redshift data (orange). Top:
Distance modulus ($\mu$) from binned data (black bars) and for each
SN (red, orange circles). The dashed gray line shows the best-fit model, while the green and
blue dotted lines show models with no dark energy and matter densities $\Omega_{\rm m} = 0.3$ and $1.0$
respectively. Bottom: Residuals to the best fit model; the error bars show 68 \% confidence.
Credit: Dark Energy Collaboration \cite{DES_SN}).
}
\label{DES_Y3_SN}
\end{figure}
%---------------------

%-------------------
\begin{figure}
\includegraphics[width=0.80\textwidth]{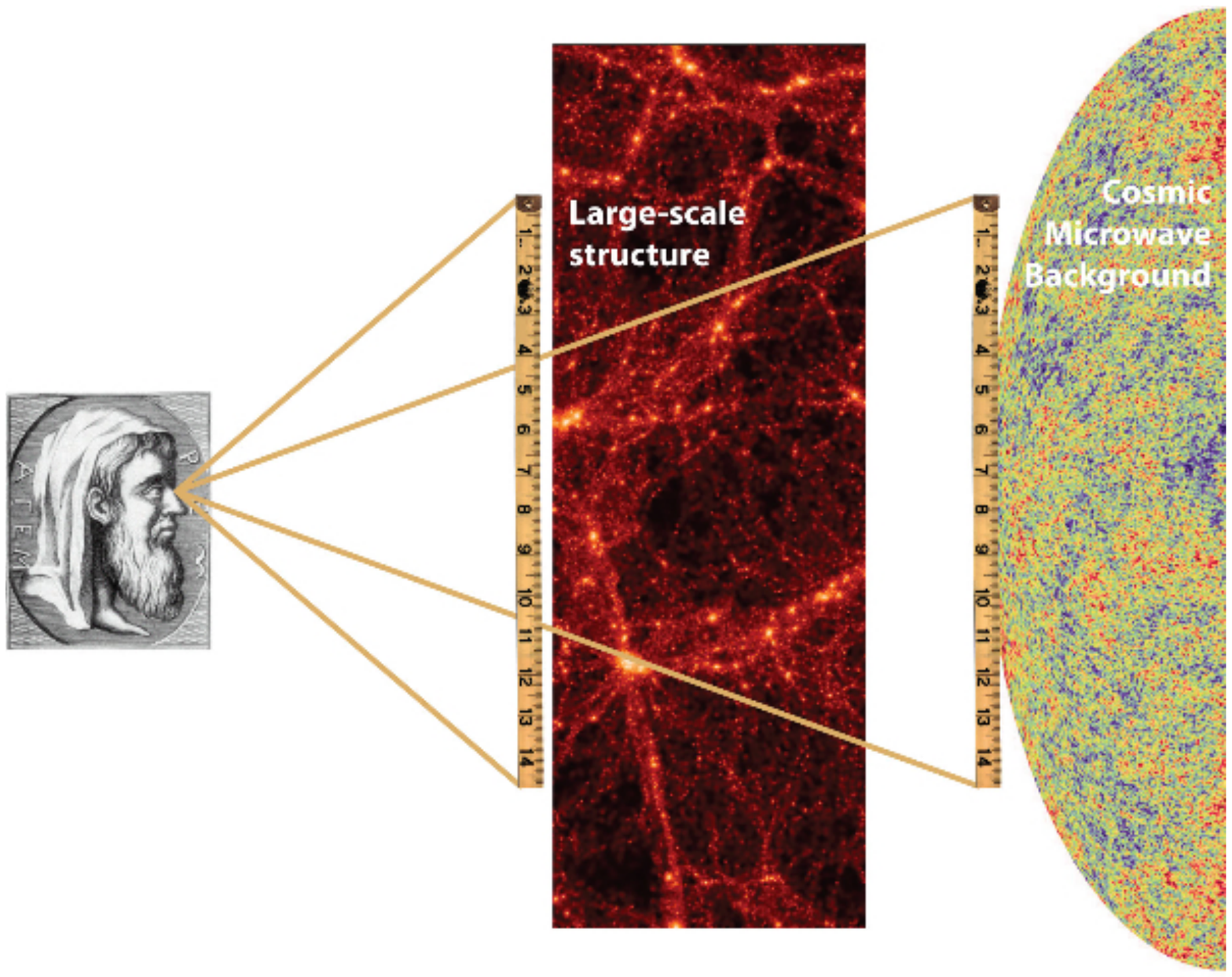}
\caption{Baryonic Acoustic Oscillations (BAO): Galaxy clustering as a probe of the geometry of the universe. The same acoustic features (BAO) seen in the CMB can be observed in the distribution of galaxies, providing a standard cosmological ruler. Credit: Euclid collaboration.
}
\label{BAO}
\end{figure}
%---------------------

Observational Cosmology provides tests of Einstein's General Relativity and alternative theories. The subject has developed in recent decades from `metaphysics'  to a highly quantitative discipline, as recognised by the award of the Nobel Prize in Physics 2019 to Jim Peebles.
Among the well-established cosmological probes are: Type Ia Supernovae, large-scale galaxy clustering, galaxy clusters, and weak gravitational lensing. 
Since the methods are sensitive to cosmological parameters in different ways, we can use them in combination to maximize our knowledge, and to control systematic errors. 

%------------------
\subsection{Supernovae Type Ia as standard candles}
Exploding stars called Supernovae Type Ia are understood to originate in binary systems in which at least one of the two companion stars is a white dwarf. 
A Supernova becomes as bright as an entire galaxy of tens of billions of stars and then begins to fade away within a matter of weeks.  
Supernovae  Type Ia are observed to all have nearly the same absolute luminosity when they reach their peak light (after some calibration corrections), hence the concept of `standard candles'.  By comparing the apparent brightnesses of different Supernovae we can determine their relative distances. We can also measure the redshift to the Supernova (or its host galaxy), and look at the relation of distance and redshift (the so-called Hubble diagram). This relation depends on the geometry of the Universe, and hence on the amounts of  Dark Matter and Dark Energy. This  method led to the discovery that the Universe is accelerating at present \cite{Perl1999, Riess1998},  which was recognised by the award of the Nobel Prize in Physics 2011. See Figure \ref{DES_Y3_SN} for recent results. 

%---------------------
\subsection{Baryonic acoustic oscillations as standard rulers}
Another major cosmological probe is the spatial clustering of galaxies (deviation from a random uniform spatial distribution). What is useful as a standard ruler is  a feature  called baryon acoustic oscillations (BAO), first discovered by the 2dF and SDSS teams {\cite{cole2005, Eisenstein2005}. Until approximately 370,000 years after the Big Bang, the Universe was a sea of  colliding protons, electrons, and photons, forming an opaque, ionized plasma. Gravity pulled the protons and electrons together, while the pressure of the energetic photons  acted to push them apart. This competition between gravity and pressure created a series of sound waves, `acoustic oscillations', in the plasma. As the Universe expanded and cooled, the photons lost energy and became less effective at ionizing atoms, until gradually all the protons and electrons combined to form neutral hydrogen. Thereafter, the Universe became transparent to photons, which travelled without further interacting with matter. The distance travelled by the sound waves up to that point became imprinted in the spatial distribution of matter, which means that today there is a slight tendency for pairs of galaxies to be separated by about 480 million light years. Comparing this scale to how much galaxies \textit{appear} to be separated on the sky, i.e., their angular separation, allows us to work out how far away they are. Comparing this distance to the redshift of their light, just as we do for Supernovae, tells us about the expansion history of the Universe. 
The same acoustic features (BAO) are seen in the CMB. See an illustration in Figure \ref{BAO}.
There is  additional useful information in galaxy clustering, in particular in  `redshift distortion' (the deviation of galaxy motions from the smooth Hubble flow) and in the entire statistical fluctuations at different physical scales.
There are other important probes of the mass distribution, in particular `Lyman-$\alpha$ clouds' (inferred from spectra of Quasars) 
and peculiar velocities (deviations from the smooth Hubble flow).

%------------------------
\subsection{Cluster abundance}
Galaxy clusters, first catalogued by George Abell and others, are peaks in the `cosmic web' of the galaxy distribution.
Examples of galaxy clusters are shown in two insets in Figure  \ref{DES_WL_map}.
As the geometry  and growth of structure depend on Dark Energy,  counting the number of galaxy clusters of a given mass within a given volume of the Universe 
provides another cosmological test. The main challenge when interpreting cluster counts is that the mass of a cluster (which is mostly Dark Matter) is not directly observable, so we have to calibrate mass-observable relations using techniques such as the Sunyaev-Zel'dovich effect, X-ray emission, and calibration by weak gravitational lensing. Powerful computer simulations play a key role in understanding the formation of structure and in assessing the accuracy of these cluster mass estimates.

%-------------------
\begin{figure}
\includegraphics[width=0.80\textwidth]{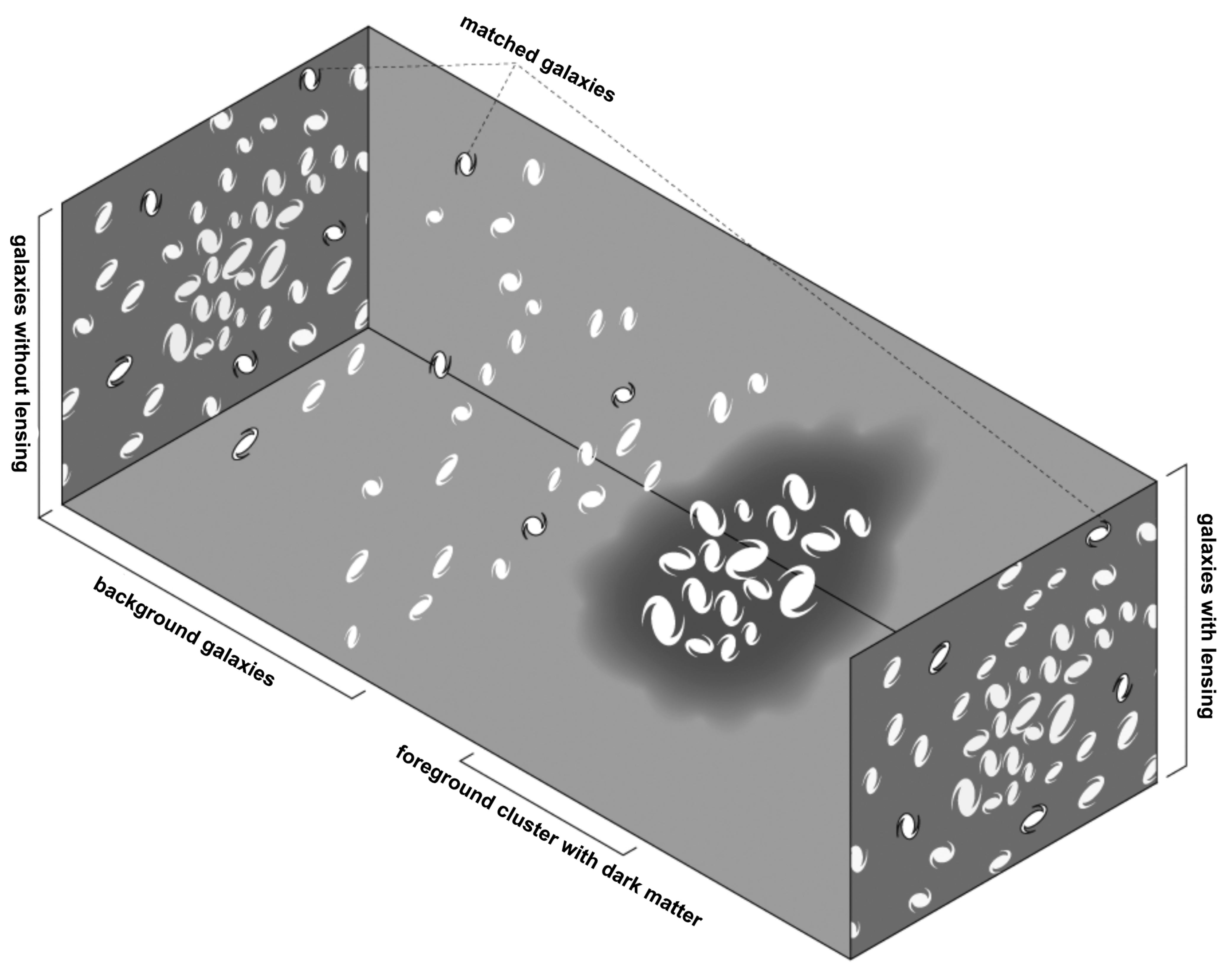}
\caption{Gravitational Lensing:
A schematic diagram showing the effects of gravitational lensing. The observer (on the right hand side)  would see distorted images by the intervening 
clumpy distribution of  dark matter. 
Credit: Michael Sachs under Creative Commons.

}
\label{WL_diagram}
\end{figure}
%---------------------

%-------------------
% From a article DOE, based on Vikram, Chang et al.
%http://deswl.github.io/page1/vikram_paper/vikram_paper.html
% https://www.newswise.com/doescience/?article_id=632609&returnurl=aHR0cHM6Ly93d3cubmV3c3dpc2UuY29tL2FydGljbGVzL2xpc3Q=
\begin{figure}
\includegraphics[width=0.80\textwidth]{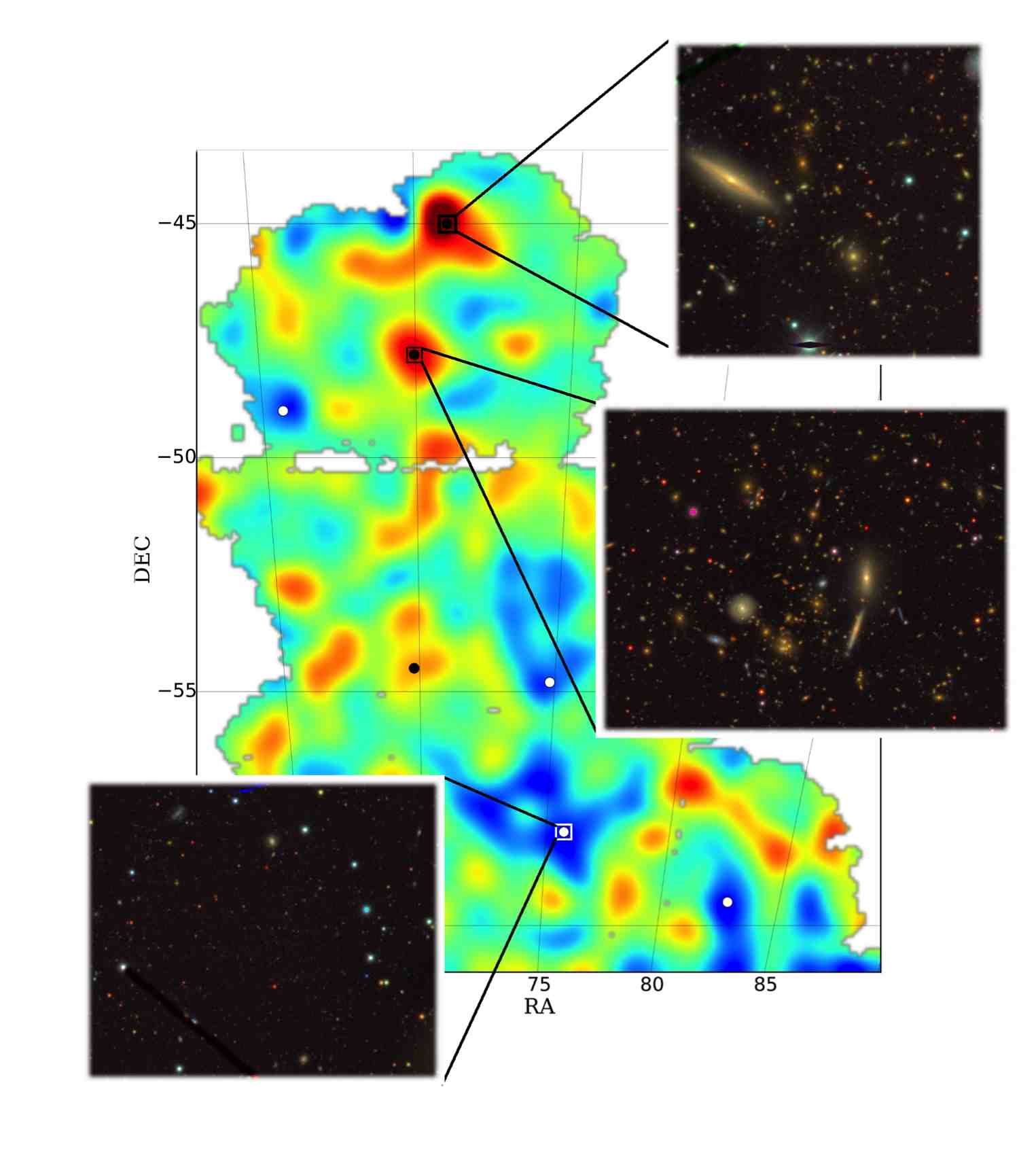}
\caption{
The distribution of dark matter projected on the sky  (across 139 square degrees). The dark mass map was derived from weak gravitational lensing by the DES collaboration.  The colour scale represents the density of mass: yellow and red are regions with more dense matter; green and blue are regions with less dense matter.  As examples of the correspondence of mass to light, two regions (in red) of dark matter over-densities are identified with known galaxy clusters, while an under-dense region (in blue) corresponds to a void.
Credit: The Dark Energy Survey collaboration (based on \cite{Chang2015}).
}
\label{DES_WL_map}
\end{figure}
%---------------------

\subsection{Weak gravitational lensing}
Gravitational lensing was predicted by Einstein's General Theory of Relativity. 
The effect is that light rays are bent when they pass through a gravitational field. 
This was first proved by the solar eclipse experiment in 1919 by Eddington and Dyson. 
Bending of light from a distant galaxy by a massive cluster along the line-of-sight generates `strong lensing', with the distorted images in the form of apparent `arcs'. 
Milder intervening mass fluctuations produce `weak lensing' (see an illustration in Figure \ref{WL_diagram}).
Weak gravitational lensing is sensitive to both the expansion history and the growth history of density fluctuations, and therefore is sensitive to the amounts of Dark Matter and Dark Energy.  
 The apparent distortions of distant galaxies in a small patch of sky will be correlated, since their light has travelled through nearly the same intervening field of Dark Matter; we can use this correlation to statistically tease out the small lensing distortion. This effect, known as cosmic shear, is a tiny distortion of about one percent on average, and it requires very precise, accurately calibrated measurements of many galaxy shapes for its inference. 
 The distorted images can be used to infer a projected mass map (see Figure \ref{DES_WL_map}). The growth of clumpiness with cosmic time tells us about the 
 amount of Dark Energy.

\subsection{The speed of Gravitational Waves }
A somewhat unexpected probe of  Dark Energy has emerged from the  discovery of Gravitational Waves. The first event, GW150914, a Binary Black Hole merger, was detected by LIGO \cite{LIGObbh} on 14 September 2015, and subsequently was recognised by the award of the Nobel Prize in Physics 2017.  The later he LIGO \& Virgo detection \cite{LIGObns} of Binary Neutron Star (BNS) merger GW170817 and electromagnetic followups turned out to be relevant for constraining cosmological models. Several ground and space telescopes (including the Dark Energy camera \cite{des-gw1} 
in Chile) captured an optical flash from this Gravitational Wave event.
In particular, the Fermi Gamma-ray space telescope  detected a flash from the BNS system only 1.7 seconds  after the  Gravitational Wave measurement.
Given the distance of about 130 Million Light years to the host galaxy NGC4993, this implies that the speed of Gravitational Waves  is equal to the speed of light to within  one part in $10^{15}$.  This has ruled out certain models of gravity, which have predicted, at least in their simplest form, differences between the speed of gravity and light.

%-----------------------------------------------------------------------------------------------------------------------------------------
\section{The landscape of galaxy surveys}

%-----------------------
\begin{figure}
\includegraphics[width=1.0\textwidth]{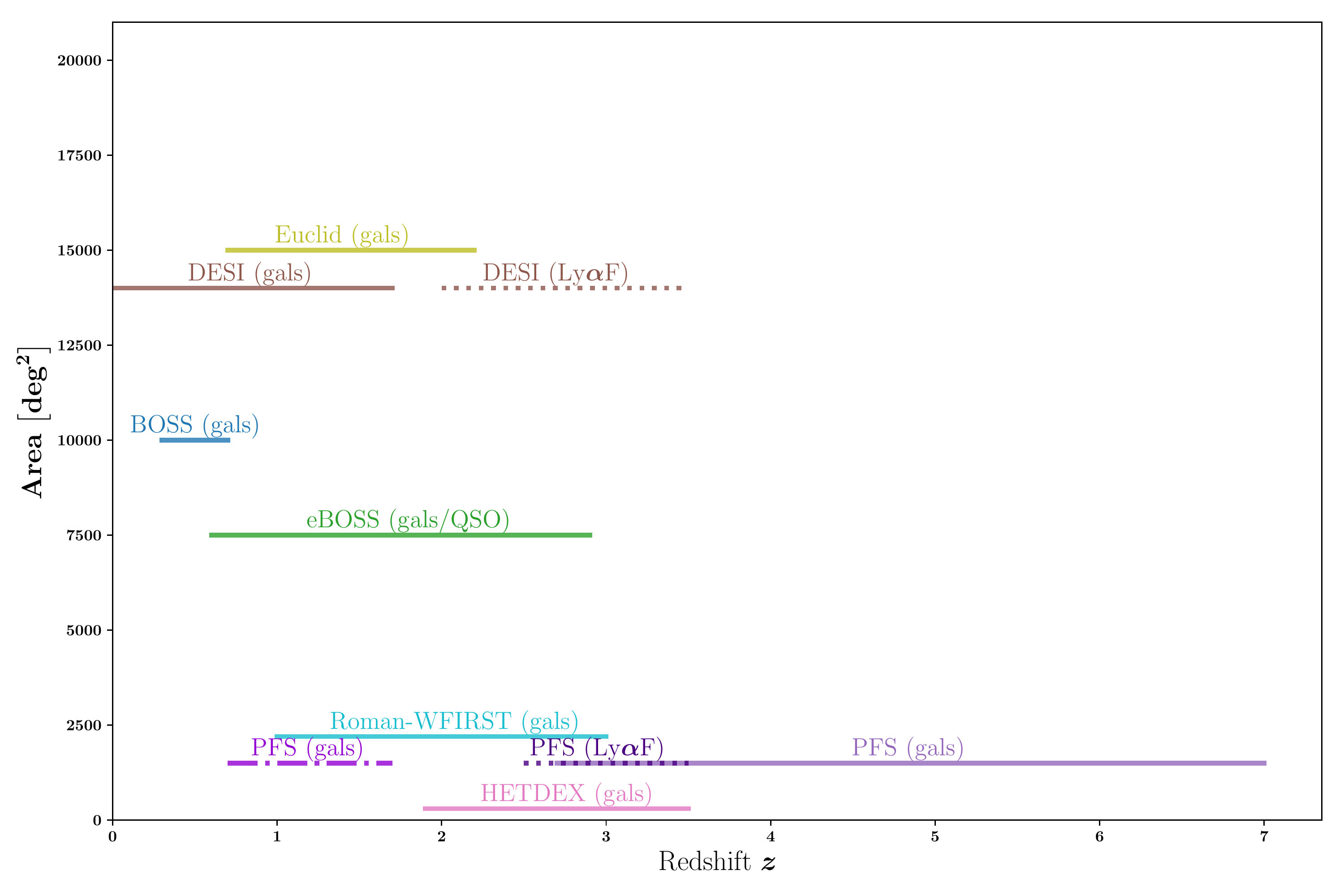}\
\caption{The survey area vs. redshift ranges for a selection of current and spectroscopic surveys.
`gals'  and `QSO' indicate surveyed galaxies and quasars. 'Ly$\alpha$F' stands for Lyman-$\alpha$ Forest, 
a technique that provides the distribution of gas clouds along the line-of-sight to distant quasars.
Credit: Krishna Naidoo's PhD thesis (UCL), based on a Table given in  \cite{Weinberg2019}. }
\label{spec_surveys}
\end{figure}
%--------------------------

%-----------------------
\begin{figure}
\includegraphics[width=1.0\textwidth]{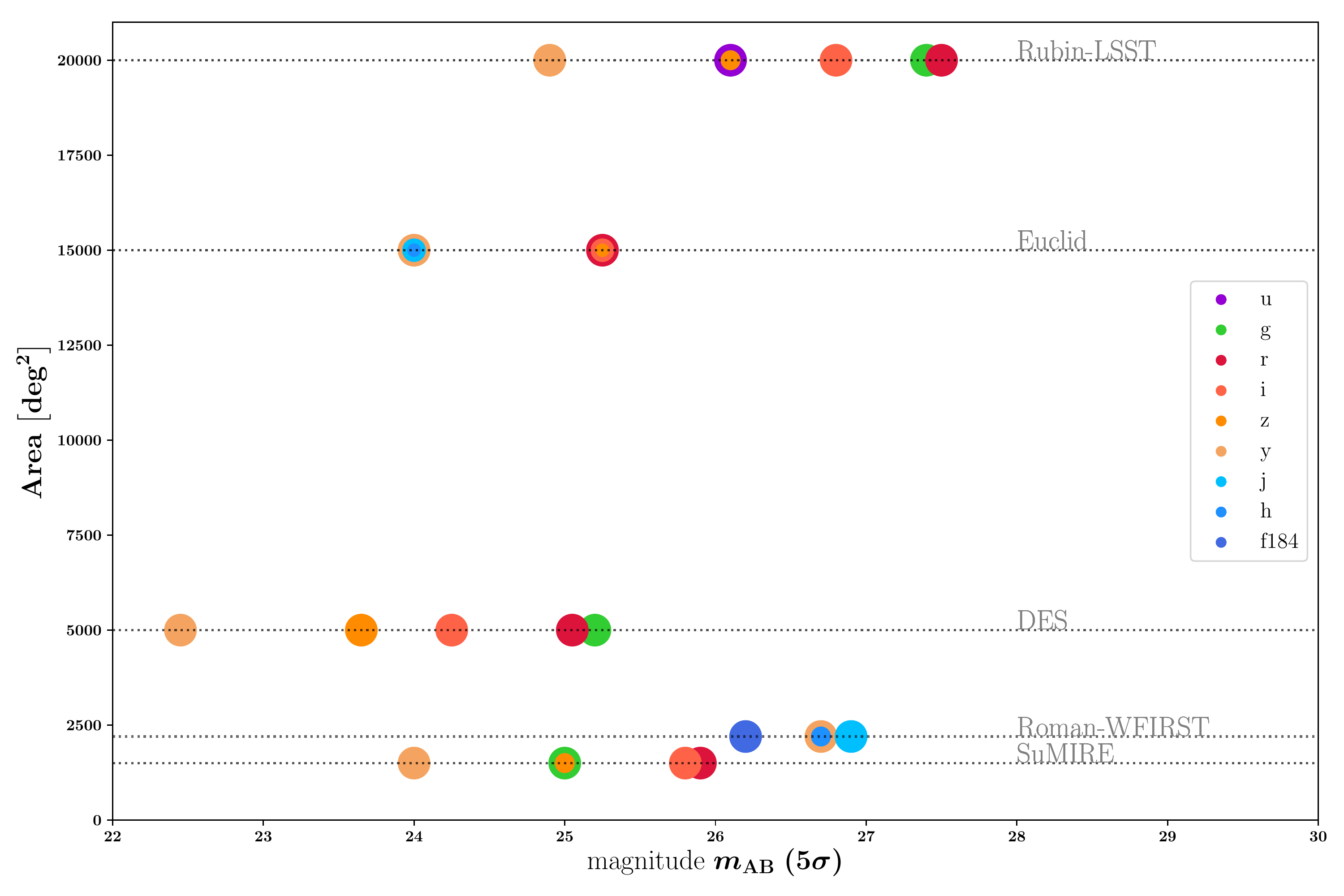}\
\caption{The survey area vs. limiting apparent magnitudes (per filter) for a selection of imaging surveys. The larger the magnitude, the fainter the object.
The (monochromatic)  AB magnitude system is defined as $m_{\rm AB} =-2.5 \log_{10}(f_{\nu}) + 8.90$, where the spectral flux density$f_{\nu}$ is in Janskys. The $5 \sigma$ indicates the level of signal-to-noise.
Credit: as in Figure \ref{spec_surveys}.}
\label{photo_surveys}
\end{figure}
%--------------------------

The 1998-1999 results of the accelerating Universe from the Supernovae Type Ia observations have stimulated many galaxy surveys designed to verify and characterise Dark Energy.
Back in 2006, the U.S. Dark Energy Task Force (DETF) report 
%(\citealt{2006astro.ph..9591A}) 
classified Dark Energy surveys into numbered stages: Stage II projects were on-going at that time; Stage III were near-future, intermediate-scale projects; and Stage IV were larger-scale projects in the longer-term future. These projects can be further divided into ground-based and space-based surveys. Space-based missions have the advantage of not having to observe through the Earth's atmosphere, which blurs the light and absorbs infrared light.
Further, surveying in Astronomy is divided into spectroscopic and imaging. 
Spectroscopic surveys produce accurate redshifts of galaxies, but are demanding in terms of observing time. Multi-band photometric surveys 
can measure far more galaxies, but the deduced `photometric redshifts' are less accurate than spectroscopic redshifts.
Among the spectroscopic surveys we note:
the current SDSS Baryon Oscillation Spectroscopic Survey (BOSS)\footnote{http://www.sdss3.org/surveys/boss.php}, eBOSS (`extended BOSS')\footnote {https://www.sdss.org/surveys/eboss/},  the Dark Energy Spectroscopic Instrument (DESI)\footnote{{http://desi.lbl.gov/}} and under construction  
the Subaru Prime Focus Spectrograph (PFS)\footnote{{http://pfs.ipmu.jp/factsheet/}}, 4MOST\footnote{{http://www.4most.eu/}}, HETDEX\footnote{https://hetdex.org/hetdex.html}, 
Euclid\footnote{{http://www.euclid-ec.org/}} and the Nancy Roman Wide-Field Infrared Survey Telescope (WFIRST)\footnote{http://www.stsci.edu/wfirst}.
The sky area vs. redshift range are shown for some of these surveys in Figure \ref{spec_surveys}.

Current imaging surveys include 
the Dark Energy Survey (DES)\footnote{http://www.darkenergysurvey.org/},
the  Hyper Suprime Cam (HSC)\footnote{{http://www.naoj.org/Projects/HSC/}}, the Kilo-Degree Survey (KiDS)\footnote{{http://kids.strw.leidenuniv.nl/}}, PAU\footnote{https://www.pausurvey.org}, and under construction 
the Vera Rubin Observatory of Legacy Survey of Space and Time (LSST)\footnote{{http://www.lsst.org/}}, and the above-mentioned Euclid and Roman-WFIRST. Figure  \ref{photo_surveys} shows the sky area vs. magnitudes (per filter) for some of these imaging surveys.

%-----------------------
\begin{figure}
\includegraphics[width=0.30\textwidth]{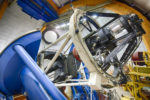}\
\includegraphics[width=0.30\textwidth]{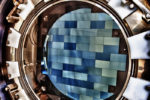}\
\includegraphics[width=0.30\textwidth]{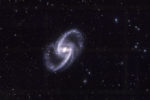}
\caption{From left to right: the Dark Energy Camera (DECam) on the Blanco 4m telescope in Chile, the CCD array in the focal plane,  and the galaxy NGC1365,  a barred spiral galaxy 56 million light-years away,  observed with DECam at `first light' in 2012. Credit: the DES collaboration}
\label{DES_Blanco}
\end{figure}
%--------------------------

\section{The Dark Energy Survey}
\label{subsec:2}

As discussed in the previous Section, many ongoing and planned imaging and spectroscopic surveys  aim at measuring Dark Energy and other cosmological parameters. As a showcase  we  focus here on DES\footnote{I have chosen  DES as a showcase  as its observations are already complete. 
Also, I happen to have an 'insider's view', as I have been 
involved in the project since its early days back in 2004, in particular as co-chair of its Science Committee until 2016, and later as chair of its Advisory Board.}. For an overview of  DES see the article `DES: more than Dark Energy' \cite{DES2016} and the DES book \cite{DES_book_2020}, which tells the story of this experiment, from construction to science.

In short, DES is an imaging survey of 5000 square degrees of the Southern sky, utilising a 570 mega-pixel camera (called DECam) on the 4m Blanco telescope in Chile (see Figure \ref{DES_Blanco}). Photometric redshifts (approximate distances) to the galaxies were obtained from the multi-band photometry to produce a three dimensional map of 300 million galaxies. The main goal of DES is to determine the Dark Energy equation of state $w$ and other key cosmological parameters to high precision. DES has measured $w$ using four complementary techniques (described in Section 3) in a single survey: galaxy clustering, counts of galaxy clusters, weak gravitational lensing and thousands of Type Ia Supernovae in a `time domain' survey over 27 square degrees. DES is an international collaboration, with more than 400 scientists from 26 institutions in the US, the UK, Spain, Brazil, Germany, Switzerland and Australia involved. 
The survey had its first light in September 2012 and started observations in August 2013. Observations were completed in 2019, over 758 nights spread over 6 years.

DES  imaged about an eighth of the sky to a depth of approximately 24th magnitude 
(about 15.8 million times fainter than the dimmest star that can be seen with the naked eye). The quality of the data is excellent -- the typical blurring of the images due to the camera, the telescope optics, and the atmosphere is about 0.9 arcseconds. On a typical observing night, the camera took about 200 images;  since each DECam digital image is a gigabyte in size, DES typically collected about 200 gigabytes of data on clear nights, `big data' for an astronomy experiment. 
In addition to serving the scientific interests of the collaboration, the DES 
data processed by DESDM are being made publicly available on a regular basis and can be downloaded from the internet by anyone in the world. 
DES scientists are working closely with data sets and scientists from other surveys and observatories.
Comparing DES data with complementary data from other surveys allows us to obtain new information and make new discoveries. 

\begin{figure}
\centering
\subfloat[Constraints on cosmological parameters determined by DES Year 1 galaxy clustering and weak lensing (blue), Planck CMB measurements (green), and the combination of the two (red), assuming the $\Lambda$CDM model. Within the measurements' accuracy, the Planck and DES constraints are consistent with each other.
Here, $\Omega_{\rm m}$ is the matter density parameter, and $S_8$ is a parameter related to the amplitude of density fluctuations (see text). For each colour, the contour plots represent 68\% and 95\% confidence levels. Credit: \cite{DES2018}. ]{%
\resizebox*{5cm}{!}{\includegraphics{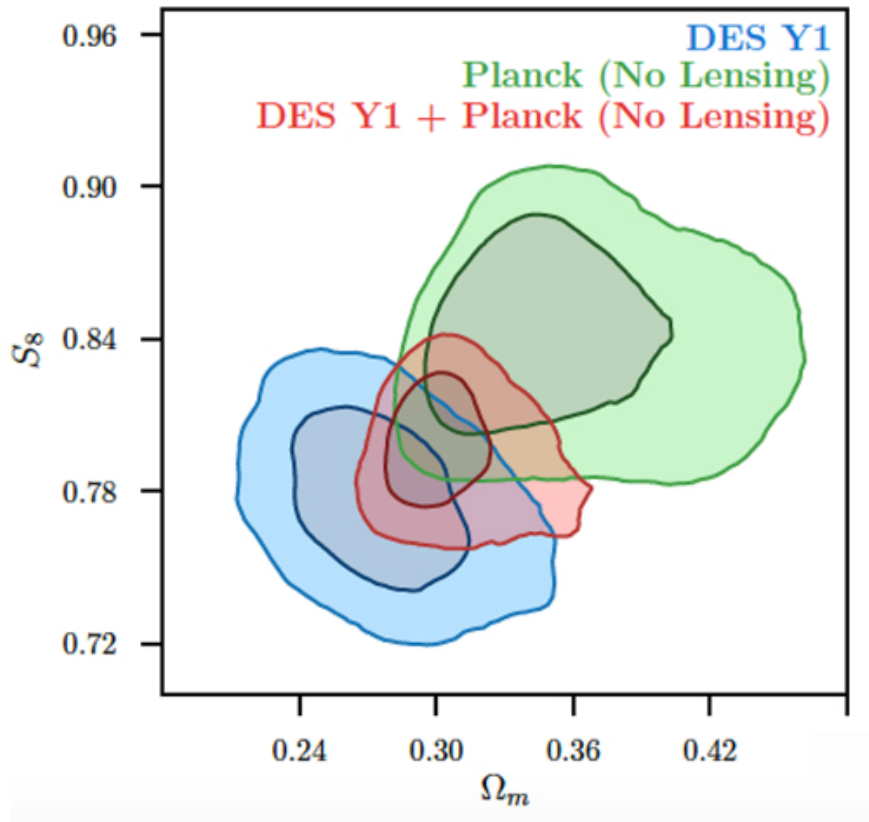}}}\hspace{5pt}
%}
\subfloat[Marginalised posterior contours in the  $\Omega_m$-$S_8$ plane. The three Weak Lensing experiments (KiDS, DESY1, HSC) tend to give a clumpiness $S_8$ lower than that from the CMB Planck data. Credit: \cite{KIDS2018}.]{%
 \resizebox*{5cm}{!}{\includegraphics{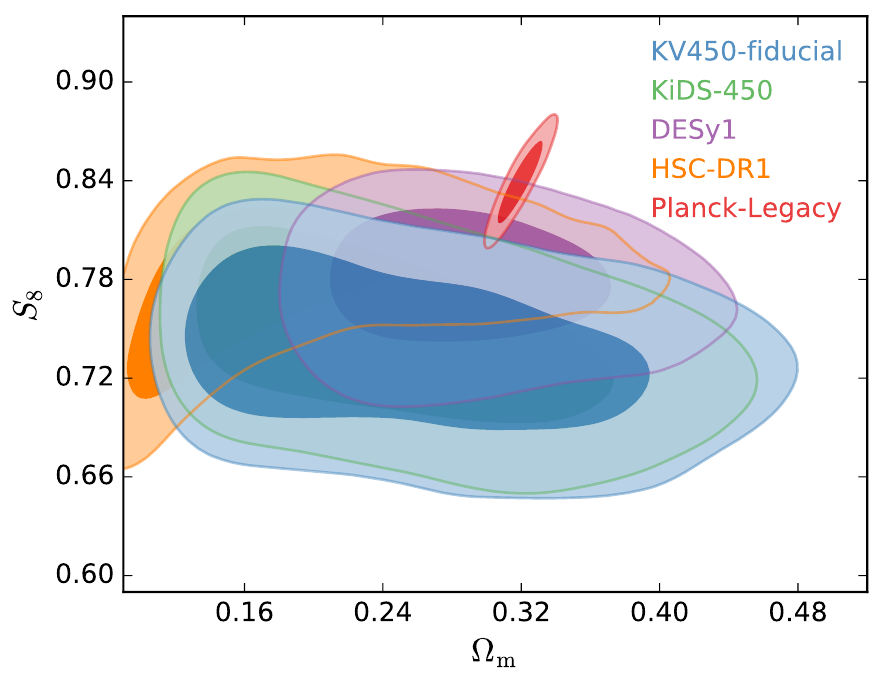}}}
\caption{Examples of `tension' in the standard $\Lambda$CDM model.}
 \label{tension-figure}
\end{figure}

\subsection{DES joint galaxy clustering and weak lensing: Year One cosmology results}

The first major DES cosmology results were announced in August 2017, and published the following year (\cite{DES2018}). 
This analysis combined galaxy clustering and weak gravitational lensing data from the first year of DES survey data that covered 
about 1300 square degrees, 
%1321 sq deg, 
utilising measurements of three two-point correlation functions (hence referred to as `3 times 2pt'): 
(i) the cosmic shear correlation function of 26 million source galaxy shapes in four redshift bins, (ii) the galaxy angular auto-correlation function of 650,000 luminous red galaxy positions in five redshift bins, and (iii) the galaxy-shear cross-correlation of luminous red galaxy positions and source galaxy shears.
The headline results strongly support the $\Lambda$CDM cosmological model. Combining
the DES Year 1 data with Planck CMB measurements, baryonic acoustic oscillation measurements from the SDSS, 6dF, and BOSS galaxy redshift surveys, and Type Ia Supernova distances from the Joint Light-curve Analysis, this analysis finds the Dark Energy equation of state parameter to be $w=-1.00^{+0.05}_{-0.04}$ (68\% CL) , in spectacular agreement with the cosmological constant model, for which $w=-1$. 

Contrasting the DES and Planck measurements yields an important comparison of the early and late Universe. Planck measures conditions at the epoch of ``last (photon) scattering" 370,000 years after the Big Bang. 
Assuming $\Lambda$CDM, the Planck measurement predicts the amount of large-scale clumpiness in the mass distribution today. The DES measurement of that clumpiness is in reasonably good agreement with that prediction,  as shown by the overlap of the blue and green contours in Figure \ref{tension-figure} (a). This  is subject to possible parameter `tension' due to either systematics or the need for new Physics. Both possibilities are currently under further investigation.
When the DES and Planck results are combined to obtain tighter constraints, shown by the red contours, it gives  for the ratio of the matter density  to critical density $\Omega_{\rm m} = 0.298 \pm 0.007$ and for the clumpiness factor  $S_8 = \sigma_8 (\Omega_{\rm m}/0.3)^{0.5} = 0.802 \pm 0.012$ ($\sigma_8$ is the rms fluctuation in spheres of radius of
 $8 (H_0/100)^{-1}$  Mpc).
The future analysis of the full DES data (Years 1-3 and then Years 1-6), which covers a larger area of sky and goes deeper than the Year 1 data,  will provide stronger constraints on these parameters as well as on the time variation of the equation of state parameter of Dark Energy, modified gravity and neutrino mass.

\subsection{DES Supernova Year 3 cosmology results}

The first DES Supernova Ia (SNe Ia) cosmology results were announced in
January 2018  and published a year later \cite{DES2019}.  The analysis used a sample of 207 spectroscopically confirmed SNe Ia from the first three years of DES-SN, combined with a low-redshift sample of 122 SNe from the literature, thus 329 SNe Ia in total (see Figure \ref{DES_Y3_SN}).  The $\Lambda$CDM model, combining DES-SN with Planck, yields  a matter density $\Omega_{\rm m}= 0.331 \pm  0.038$; for the $w$CDM model (with constant $w$), the analysis finds 
$w = - 0.978 \pm  0.059$, and $\Omega_{\rm m} = 0.321 \pm 0.018$.  
These results strongly support the $\Lambda$CDM paradigm, and they agree with previous constraints using SNe Ia. Future DES SNe Ia analyses will use a much larger sample of several thousand photometrically classified Supernovae from the full survey.

%+++++++++++++++++++++++++++++++++++++++++++++++++++++++++++++++++++++++++++++++++++++++++++

\section{Tension in clumpiness and in  the Hubble constant: systematics or new physics?}

\begin{figure}
\includegraphics[width=\textwidth]{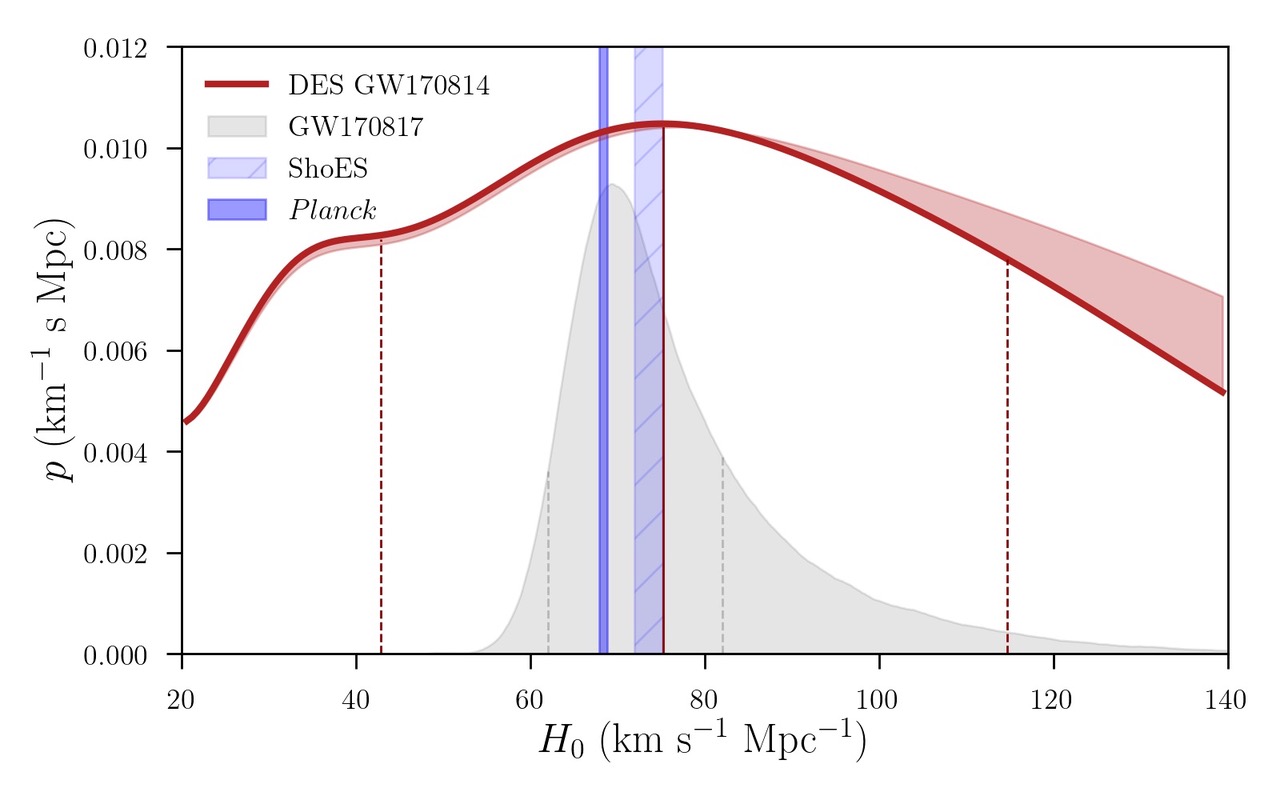}
\caption{Four methods for measuring the Hubble Constant: 
from the cosmic ladder SH0ES \cite{hubble:riess19}, from the CMB Planck \cite{hubble:planck2018VI}, 
from the Gravitational Wave event GW170817 associated with the galaxy NGC4993 as a Bright Siren \cite{hubble:bright_siren}  and from GW170814 as Dark Siren \cite{hubble:dark_siren},
where the Hubble constant posterior distribution was obtained by marginalizing over 77,000 possible host galaxies catalogues by DES.
Credit: \cite{hubble:dark_siren}}
\label{H0_dark_siren}
\end{figure}

The $\Lambda$CDM model has survived detailed observational tests over 30 years. 
However, it is still undergoing `health checks' from time to time.  $\Lambda$CDM is the winning model, as demonstrated  e.g. in Planck \cite{Planck2018I}, DES \cite{DES2018} and recently by the  CMB Atacama Cosmology Telescope (ACT) \cite{ACT2020} and eBOSS \cite{eBOSS2020} results.
But we have already indicated in the previous chapter some inconsistency in the clumpiness parameter $S_8$: 
the one derived from weak lensing is about 3-$\sigma$ smaller than the value deduced from Planck (see e.g. the recent KiDS results \cite{KIDS2020}).

An even larger tension, of about 4.4-$\sigma$, exists between the Hubble Constant measurements from the cosmic ladder in the nearby universe and from the CMB experiments  that probe the early universe.
Edwin Hubble discovered the law of expansion of the Universe
by measuring distances to nearby galaxies.  The Hubble Constant is derived from  the ratio 
of the recession velocity $v$ and the distance $d$:
\begin{equation}
H_{0} = v/d.
\end{equation}
Note $H_0$ has units of inverse time, although the units are written as ${\rm km} \, {\rm s}^{-1} \, {\rm Mpc}^{-1}$ to reflect the way it is being measured. 
Astronomers have argued for decades about the
systematic uncertainties in various methods and derived values over
the wide range $50 \, {\rm km} \, {\rm s}^{-1} \, {\rm Mpc}^{-1}\simlt
H_0 \simlt 100 \, {\rm km} \, {\rm s}^{-1} \, {\rm Mpc}^{-1}$.

One of the most reliable results on the Hubble constant came from the
Hubble Space Telescope (HST) Key Project \cite{hubble:freedman2001}.  This study
used the empirical period--luminosity relation for Cepheid variable
stars, and calibrated a number of secondary distance
indicators, SNe Ia,
the Tully--Fisher relation, surface-brightness fluctuations, and Type
II Supernovae.  This approach was further extended, based on HST 
observations of 70 long-period Cepheids in the Large Magellanic Cloud combined with  Milky Way parallaxes and masers in NGC4258, to yield  
$ H_0 = 74.0 \pm1.4 \;{\rm km} \, {\rm s}^{-1} \, {\rm Mpc}^{-1}$ \cite{hubble:riess19}\ (SH0ES).

By contrast,  the Planck CMB experiment \cite{hubble:planck2018VI} gives a lower value, 
$H_0 =67.4 \pm 0.5\;{\rm km} \, {\rm s}^{-1} \, {\rm Mpc}^{-1}$.   
The tension  of 4.4-$\sigma$ between  $H_0$ from Planck and the traditional cosmic
distance ladder methods is of great interest and under investigation.
A low value,  $H_0 = 67.9 \pm 1.5 \;{\rm km} \, {\rm s}^{-1} \, {\rm Mpc}^{-1}$, has  also been measured by ACT\cite{ACT2020}, another CMB experiment. 

Other methods have been used recently, for example by 
calibrating of the tip of the red-giant branch applied
to SNe Ia,
finding
$ H_0 = 69.8 \pm0.8 \; {\rm stat.} \; \pm 1.7 \; {\rm sys.} \;{\rm km} \, {\rm s}^{-1} \, {\rm Mpc}^{-1}$
\cite{hubble:freedman2019}.
The method of 
time delay in gravitationally-lensed quasars \cite{hubble:Holicow}\ (H0LiCOW)
gives the result 
$ H_0 = 73.3^{+1.7}_{-1.8} \;{\rm km} \, {\rm s}^{-1} \, {\rm Mpc}^{-1}$, but see a revision \cite{Holicow2020} to 
$ H_0 = 67.4^{+4.1}_{-3.2} \;{\rm km} \, {\rm s}^{-1} \, {\rm Mpc}^{-1}$.

Another method that recently came to fruition is based on Gravitational Waves; the `Bright Siren' applied to  the binary neutron star merger GW170817
and  the `Dark Siren' implemented on the binary black hole merger GW170814 yield
$ H_0 = 70^{+12}_{-8} \;{\rm km} \, {\rm s}^{-1} \, {\rm Mpc}^{-1}$ \cite{hubble:bright_siren} 
and 
$ H_0 = 75^{+40}_{-32} \;{\rm km} \, {\rm s}^{-1} \, {\rm Mpc}^{-1}$ \cite{hubble:dark_siren} respectively.
With many more gravitational wave events, the future uncertainties on $H_0$ from standard sirens will get smaller.
Figure \ref{H0_dark_siren} summarizes four of the methods discussed above. 

For further discussion of the Hubble Constant measurements see \cite{verde2019} and \cite{gpe2020}.
There is possibly a trend for higher $H_0$ at the nearby Universe and a lower $H_0$ at the early Universe, which leads some researchers to propose a time-variation of the Dark Energy component or other exotic scenarios. 
Ongoing studies are addressing the question of whether the Hubble tension is due to systematics in at least one of the probes, or is a signature of new Physics. 
On the whole, although there are inconsistencies in the Hubble Constant and clumpiness factor, they are not at the level of shaking up the $\Lambda$CDM paradigm.

\section {Discussion and Outlook}

The accelerated expansion of the Universe and the $\Lambda$CDM model have been overwhelmingly supported by a host of  cosmological measurements. 
But we still have no clue as to what is causing the acceleration, and what Dark Matter and Dark Energy are. 
Even if Dark Energy is `just'  the enigmatic Cosmological Constant $\Lambda$, we desperately need to understand it.

It may well be that the $\Lambda$CDM model is indeed the best description of our Universe, with Dark Matter and Dark Energy ingredients that will eventually detected independently, e.g. in the lab. But there is also a chance that the Dark sector is the `modern Ether' and future generations will adopt an entirely different description of the Universe. It is also possible that the community has converged on a single preferred model due to `over communication' \cite{L2002}}.
Should a discrepancy between data and the existing cosmological theory be resolved by adding new entities such as Dark Matter and Dark Energy, or by modifying the underlying theory? 
This  reminds us of two cases in the history of studies of our own Solar System. Anomalies in the orbit of Uranus were explained by hypothesizing a previously unseen (i.e. `dark') planet, Neptune, within Newtonian gravity, which was subsequently discovered.
On the other  hand, anomalies in the perihelion of Mercury were successfully explained by invoking a new theory of gravitation beyond Newton, Einstein's General Relativity, instead of by the `dark' planet Vulcan (e.g. a discussion in \cite{LM2014} and references therein). Unfortunately, history does not provide a definitive guide to choosing between dark stuff or a new theory of gravity in explaining cosmological observations.  

Various speculations have been proposed about the origins of $\Lambda$:
 e.g. that we happen to reside in an unusual patch of the universe (e.g., in a large void), mimicking the appearance of $\Lambda$;
 that the Cosmological Constant $\Lambda$ is something different to vacuum energy, as  the value predicted by fundamental theory is as
much as $10^{120}$ times larger than observations permit;  that perhaps the observed $\Lambda$ is a superposition of different contributions;  
that perhaps  modifications to  General Relativity are required;  
that a higher-level theory, connecting General Relativity to Quantum Mechanics and Thermodynamics, is still to be discovered; 
or that we should possibly consider $\Lambda$ in the context of 
a Multiverse: anthropic reasoning suggests a large number of universes (`multiverse') in which $\Lambda$ and other cosmological parameters take on all possible values. We happen to live in one of the universes that is `habitable'.

It is likely that ultimately  the full explanation of Dark Energy of the Universe may require a revolution in our understanding of fundamental Physics. 
It might even require thinking beyond the boundaries of Physics.
Michael Collins, one of the  Astronauts of the Apollo 11 mission to the Moon allegedly said ``I think a future flight should include a poet, a priest and a philosopher. We might get a much better idea of what we saw." Likewise, we might need an entirely  new perspective on the mystery of Dark Energy.

\section*{Acknowledgements}
 I express thanks to my DES and UCL collaborators for many inspiring discussions on this topic over the years.
In particular I thank Richard Ellis for initiating this article, and to Lucy Calder, Josh Frieman, Andrew Liddle, Michela Massimi, Julian Mayers and David Weinberg with whom I have interacted 
on  related review articles, including some chapters in the DES book  \cite{DES_book_2020}. Rebecca Martin, Krishna Naidoo and Lorne Whiteway have kindly commented on earlier versions of this article.
I acknowledge support from a European Research Council Advanced Grant TESTDE (FP7/291329) and STFC Consolidated Grants ST/M001334/1 
and ST/R000476/1. 
%The current CG (2018-2021) is ST/R000476/1
%The previous one (2015-2018) is  ST/M001334/1

\end{document}